\documentclass[12pt]{iopart}
\usepackage{graphicx}
\usepackage{iopams}

\begin{document}

\title{Multipartite information flow for multiple Maxwell demons}

\author{Jordan M. Horowitz$^1$}
\address{$^1$ Department of Physics, University of Massachusetts at Boston, Boston, MA 02125, USA}
\ead{Jordan.Horowitz@umb.edu}

\begin{abstract}
The second law of thermodynamics dictates the fundamental limits to the amount of energy and information  that can be exchanged between physical systems.
In this work, we extend a thermodynamic formalism describing this flow of energy and information developed for a pair of bipartite systems to many multipartite systems.
We identify a natural thermodynamic quantity that describes the information exchanged among these systems.
We then introduce and discuss a refined version.
Our results are illustrated with a model of two, competing Maxwell demons.
\end{abstract}



\maketitle


\section{Introduction}

The role of information in thermodynamics has a long history, from Maxwell's demon to Szilard's engine and on to the investigations of Bennet, Landauer, and Penrose~\cite{Leff,Maruyama2009}.
Today, there has a been a resurgence in studying this old theme spurred on by both experimental and theoretical progress in our understanding of small fluctuating systems.
In the laboratory, increased control over small systems has led to the experimental verification of fundamental ideas~\cite{Toyabe2010,Pekola2013}, such as Landauer's principle~\cite{Berut2011,Jun2014}.
Concurrently, the development of stochastic thermodynamics has provided a robust theoretical framework that can naturally include information~\cite{Sekimoto,Jarzynski2011,Seifert2012}.

This progress has lead to a number of systematic methods to incorporate information into the thermodynamics side-by-side with energy and entropy~\cite{Kim2007,Touchette2004,Sagawa2008,Horowitz2010,Sagawa2011b,Munakata2012,Mandal2012,Munakata2013,Ito2013, Deffner2013,Barato2014,Barato2014c,Sandberg2014}.
For the special class of bipartite systems, the information flow specifically has been identified as a useful thermodynamic measure of information~\cite{Sagawa2012,Sagawa2013b,Allahverdyan2009,Hartich2014,Horowitz2014}.
A pair of thermodynamic systems, say $X$ and $Y$, is bipartite if the fluctuations in  each system are independent.
When the dynamics is modeled as a Markov jump process, then the two systems never jump simultaneously, only one at a time; whereas for diffusion processes, each system experiences independent, uncorrelated noise.
For such bipartite systems, the second law of thermodynamics can be split into two distinct, positive contributions, leading to an (irreversible) entropy production for each system individually~\cite{Allahverdyan2009,Hartich2014,Horowitz2014,Horowitz2014b},
\begin{eqnarray}\label{eq:biInfoFlow}
\eqalign{
{\dot S}^X_{\rm i} = d_tS(X) + {\dot S}^X_{\rm res} -{\dot I}^X \ge 0 \\
{\dot S}^Y_{\rm i} = d_t S(Y) + {\dot S}^Y_{\rm res} - {\dot I}^Y \ge 0.}
\end{eqnarray}
Each half contains the customary contributions from the time variation of each systems' Shannon entropy, $d_tS(X)$ and $d_tS(Y)$, and the rate of reversible entropy exchange with each of their environmental reservoirs, ${\dot S}^X_{\rm res}$ and ${\dot S}^Y_{\rm res}$: as an example, imagine thermal reservoirs where these contributions are proportional to heat, \emph{e.g.}, ${\dot S}^X_{\rm res} = {\dot Q}^X/T$.
The two new contributions, ${\dot I}^X$ and ${\dot I}^Y$, represent the flow of information into and out of each system and are given by the rate of change of the mutual information between $X$ and $Y$ due to their own fluctuations.
Here, the notation $d_t$ emphasizes a time derivative of a state function, as opposed to the over-dot  signifying thermodynamic flows, such as entropy or heat. 
This information-flow framework is already finding applications, as in~\cite{Barato2014b} where it was applied to study the energetic efficiency of information processing in cellular sensing and adaptation.

In this article, we expand the applicability of this formalism to $N$ multipartite systems: a collection of systems where each one experiences independent noise.
Here, the information-theoretic quantity is the information flow between each system and all the rest.
We then refine this statement by recognizing that the dynamics of any particular system is only affected by a small group of the other systems, what we call its neighbors.
As a result, only its neighbors can influence its thermodynamics, allowing us to tighten the second law with information by restricting the information flow to just neighbors.
We then conclude with an analysis of linear systems and illustrate these results with a pair of competing Maxwell demons.

\section{Dynamics of multipartite systems}
We begin our discussion with a description of the dynamics.
Our results will apply to any thermodynamically consistent Markovian dynamics~\cite{Seifert2012}, but for our narrative we consider over-damped diffusion processes, because the notation is the most concise.
Equivalent results for underdamped diffusions and Markov jump processes follow similarly.

We have in mind a collection of $N$ systems ${\bi X} = (X_1,\cdots, X_N)$, with states ${\bi x} = (x_1, \dots, x_N)$.
While we allow arbitrary interactions among the systems, we do require that each system, like $X_k$, is coupled individually to a distinct thermal reservoir with temperature $T_k$.
This splitting of the environments is the multipartite assumption and will allow us to uniquely identify the thermodynamic flows moving into and out of each system.
Since the environments are distinct, the dynamics segregate, and the Fokker-Planck equation for the time-dependent probability density $p_t({\bi x})$~\cite{VanKampen},
\begin{equation}\label{eq:fp}
d_t p_t({\bi x})=- \sum_{k=1}^N \partial_kJ_k({\bi x},t),
\end{equation}
separately includes one (probability) current for each system,
\begin{equation}\label{eq:current}
J_k({\bi x},t) = \mu_k F_k({\bi x},{\bi \lambda}_t)p_t({\bi x})-\mu_kT_k\partial_k p_t({\bi x}),
\end{equation}
with mobility $\mu_k$ and force $F_k$ depending on time-dependent, externally-controlled parameters ${\bi \lambda}_t$.
Here and throughout, we fix Boltzmann's constant to $k_{\rm B}=1$.
If the dynamics were not multipartite, then each $J_k$ would include thermal fluctuations induced by  thermal reservoirs different from the $k$-th, which would manifest itself by the inclusion of a term like $\mu_{ki}T_i\partial_ip$ for $i\neq k$ in $J_k$.
Alternatively, on the level of individual trajectories, the multipartite assumption restricts the form of the coupled Langevin equations~\cite{VanKampen} by requiring the noise driving each system to be uncorrelated.

The multipartite structure has an important consequence that we will exploit in the following as we develop a thermodynamics of interacting systems.
Namely, any linear functional of the currents, what we will call a flow, can be divided into separate contributions arising from each current individually.
Specifically, for a generic flow we have
\begin{equation}
{\mathcal A}=\sum_k {\mathcal A}^k = \sum_k \int J_k({\bi x},t) a_k({\bi x},t) {\rm d}{\bi x},
\end{equation}
which has a piece separately due to each current $\{J_k\}$.

\section{Thermodynamics of information flow}
For a collection of $N$ systems with Markovian dynamics as in \eref{eq:fp}, the second law of thermodynamics bounds the (irreversible) entropy production rate as~\cite{Seifert2012,VandenBroeck2010}
\begin{equation}\label{eq:2law}
{\dot S}_{\rm i} = d_t S({\bi X})+{\dot S}_{\rm res} \ge 0,
\end{equation}
where the first piece is the rate of change of the systems' Shannon entropy $S({\bi X})=-\int p_t({\bi x})\ln p_t({\bi x})\, {\rm d} {\bi x}$,
\begin{equation}\label{eq:entTot}
d_t S({\bi X}) =\sum_{k=1}^N {\dot S}^k({\bi X})=  -\sum_{k=1}^N\int J_k({\bi x},t)\partial_k\ln p_t({\bi x})\, {\rm d} {\bi x}
\end{equation}
and the second term is the the entropy flow into the environmental reservoirs,
\begin{equation}\label{eq:resTot}
{\dot S}_{\rm res} = \sum_{k=1}^N \frac{{\dot Q}^k}{T_k}=\sum_{k=1}^N\frac{1}{T_k}\int J_k({\bi x},t)F_k({\bi x},\lambda_t)\, {\rm d} {\bi x},
\end{equation}
due to the heat flow into the each reservoir ${\dot Q}^k$~\cite{Seifert2012,VandenBroeck2010}.
Summing \eref{eq:entTot} and \eref{eq:resTot}, gives a compact expression for the entropy production
\begin{equation}\label{eq:entProd}
{\dot S}_{\rm i} = \sum_{k=1}^N {\dot S}^k_{\rm i}  = \sum_{k=1}^N\frac{1}{\mu_kT_k}\int \frac{J_k^2({\bi x},t)}{p_t({\bi x})}\, {\rm d}{\bi x}.
\end{equation}

Importantly, each of the quantities entering the second law is a flow, with each current contributing separately, due  to the multipartite structure.
The result is that the second law holds for each system individually; specifically for $X_k$ we have
\begin{equation}\label{eq:entK1}
{\dot S}_{\rm i}^k = {\dot S}^k({\bi X})+\frac{{\dot Q}^k}{T_k}\ge 0,
\end{equation}
with each term representing the contribution to the entropy production coming from the $k$-th environment's thermal perturbations of $X_k$.
We can refine this statement slightly, by recognizing that the environmental perturbations due to the $k$-th environment only affect $X_k$ directly and all the other systems ${\bi X}_{-k}=(\cdots,X_{k-1},X_{k+1},\cdots)$ are only disturbed indirectly.
As a result, ${\dot S}^k({\bi X})$ equals the variation in the conditional entropy $S(X_k|{\bi X}_{-k})=-\int p_t({\bi x})\ln p_t(x_k|{\bi x}_{-k})\, {\rm d}{\bi x}$ of $X_k$ given ${\bi X}_{-k}$:
\begin{equation}
{\dot S}^k({\bi X}) =  \int J_k({\bi x},t)\partial_k\ln p_t(x_k|{\bi x}_{-k})\, {\rm d} {\bi x} = {\dot S}^k(X_k|{\bi X}_{-k}),
\end{equation}
where $p(x_k|{\bi x}_{-k})=p({\bi x})/p({\bi x}_{-k})$.
Thus, the second law for $X_k$ takes the form
\begin{equation}\label{eq:entK2}
{\dot S}_{\rm i}^k = {\dot S}^k(X_k|{\bi X}_{-k})+\frac{{\dot Q}^k}{T_k}\ge 0.
\end{equation}

Next, to generalize \eref{eq:biInfoFlow}, we recognize that ${\dot S}^k(X_k|{\bi X}_{-k})$ contains in it the variation in the correlations between $X_k$ and all the other systems ${\bi X}_{-k}$.
Our goal is to make these correlations explicit.
To this end, we use the mutual information, defined for any pair of random variables $U$ and $V$ with joint probability density $p(u,v)$ as~\cite{Cover}
\begin{eqnarray}\label{eq:mutInfo}
\eqalign{
I(U;V) &= S(U) - S(U|V) = S(V) - S(V|U) \\
& = \int p(u,v)\ln\left[\frac{p(u,v)}{p(u)p(v)}\right]\, {\rm d} u{\rm d} v,}
\end{eqnarray}
in $k_{\rm B}=1$ units.
$I$ is symmetric and always positive, $I\ge 0$, being zero only when $U$ and $V$ are independent. As such, it is a information-theoretic measure of correlations.

To incorporate the mutual information into our second law for $X_k$ in \eref{eq:entK2}, we start by noting that the mutual information's time derivative is a flow, $d_tI = \sum_k{\dot I}^k$, and that its relationship with the entropy \eref{eq:mutInfo} holds on the level of flows.
In particular, for the mutual information between $X_k$ and ${\bi X}_{-k}$ we have
\begin{equation}\label{eq:infoFlow}
{\dot I}^k(X_k;{\bi X}_{-k}) = d_t{S}(X_k)-{\dot S}^k(X_k|{\bi X}_{-k}) = \int J_k({\bi x},t)\partial_k\ln p_t({\bi x}_{-k}|x_k)\, {\rm d}{\bi x}.
\end{equation}
Here, it is important to recognize that the total time derivative of $S(X_k)$ is equal to its flow in the $k$-th direction, since it can only change due to the environmental fluctuations of the $k$-th reservoir: $d_tS(X_k)={\dot S}^k(X_k)$.
This information flow quantifies how $X_k$'s fluctuations affect its correlations with the other systems.
For example, ${\dot I}^k>0$ signifies that $X_k$ is gathering information and learning about the other systems, since $I(X_k;{\bi X}_{-k})$ is increasing on average as $X_k$ evolves.
With this identification, \eref{eq:entK2} can readily be rewritten as
\begin{equation}\label{eq:2lawK}
{\dot S}_{\rm i}^k = d_t S(X_k)+\frac{{\dot Q}^k}{T_k}-{\dot I}^k(X_k;{\bi X}_{-k})\ge 0.
\end{equation}
This expression describes the information flow's effect on the entropy balance of $X_k$.
To get a feel for its implications, imagine for the moment that $X_k$ were alone.
In this case, we would identify $\sigma^k = d_t S(X_k)+{\dot Q}^k/T_k$ as its entropy production [cf.~\eref{eq:2law}], which the second law would require to be positive.
However, the presence of the extra systems allows for the possibility that the entropy of $X_k$ and its environment decreases, $\sigma^k<0$.
In other words, when ${\dot I}^k<0$ the information flow acts as a resource, allowing $X_k$ to perform otherwise impossible tasks, like cyclically extracting heat from its reservoir and turning it into work, in seeming violation of the Kelvin-Planck statement of the second law~\cite{Callen}.


\section{Refined information flow}

\begin{figure}
\centering
\includegraphics[scale=0.75]{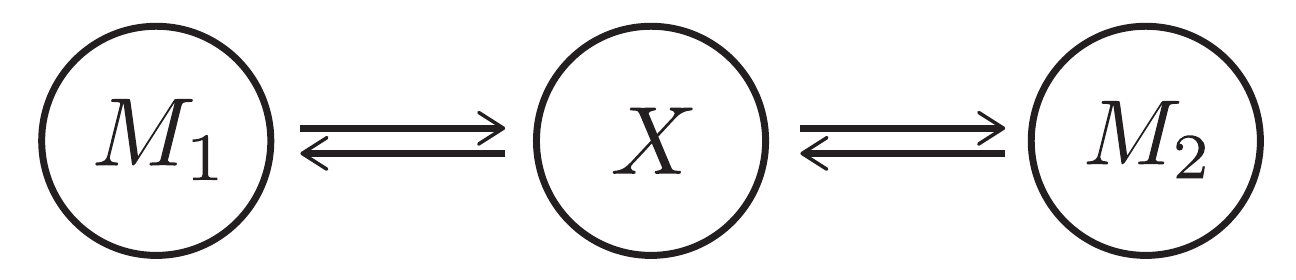}
\caption{Illustration of neighbors: Three systems $X$, $M_1$ and $M_2$ influence each other's dynamics inducing forces on each other pictured as directed arrows. $X$ has two neighbors influencing its dynamics, while the unique neighbor of both $M_1$ and $M_2$ is  $X$.}\label{fig:neighbors}
\end{figure}

We can refine \eref{eq:2lawK} by recognizing that $X_k$  can only directly learn about those systems with which it interacts -- those that directly influence its dynamics --  and that any information it has about the other systems comes indirectly.
To make this notion precise, imagine that the force on $X_k$ only depends on a subset of all the systems ${\bi X}_{\Omega_k}\subset {\bi X}_{-k}$, its neighbors, \emph{i.e.}, $ F_k=F_k(x_k,{\bi x}_{\Omega_k},\bi\lambda_t)$.
As a result, the time-variation of $X_k$ at any moment only depends on the states of its neighbors.
This idea is schematically illustrated for a tripartite system in fig.~\ref{fig:neighbors}.
With this division in mind, we can split the information flow, using the definition of mutual information (\ref{eq:mutInfo}), as
${\dot I}^k(X_k;{\bi X}_{-k}) = {\dot I}^k(X_k;{\bi X}_{\Omega_k})+{\dot I}^k(X_k;{\bi X}_{-\Omega_k}|{\bi X}_{\Omega_k}) $ into an information flow with just $X_k$'s neighbors ${\dot I}^k(X_k;{\bi X}_{\Omega_k})$ and an information flow with all the remaining systems ${\bi X}_{-\Omega_k}={\bi X}/\{X_k,{\bi X}_{\Omega_k}\}$ given $X_k$'s neighbors, ${\dot I}^k(X_k;{\bi X}_{-\Omega_k}|{\bi X}_{\Omega_k})$.
The utility of this splitting will become apparent in the following, where we show that \eref{eq:2lawK} can be split into two positive contributions,
\begin{eqnarray}\label{eq:refine1}
&d_tS(X_k)+\frac{{\dot Q}^k}{T_k}-{\dot I}^k(X_k;{\bi X}_{\Omega_k})\ge 0 \\ 
\label{eq:refine2}
&-{\dot I}^k(X_k;{\bi X}_{-\Omega_k}|{\bi X}_{\Omega_k})\ge 0.
\end{eqnarray}

Together \eref{eq:refine1} and \eref{eq:refine2} give a more precise relationship between entropy (or free energy) and information than our original bound in \eref{eq:2lawK}.
Even though $X_k$ may increase its information about ${\bi X}_{-\Omega_k}$ by way of ${\bi X}_{\Omega_k}$, ${\dot I}^k(X_k;{\bi X}_{-\Omega_k})> 0$, when we exclude those indirect correlations, $X_k$ has no way to gather information about its non-neighbors.
Thus,  we must have ${\dot I}^k(X_k;{\bi X}_{-\Omega_k}|{\bi X}_{\Omega_k})\le0$.
By contrast, $X_k$ can directly learn about its neighbors, ${\dot I}^k(X_k;{\bi X}_{\Omega_k})>0$, which costs entropy $\sigma^k=d_tS(X_k)+{\dot Q}^k/T_k$.
From the opposite perspective, the only information available to perform a useful task, which would make $\sigma^k<0$, is the information between $X_k$ and its neighbors.
A similar splitting was introduced in~\cite{Sartori2014} to study the thermodynamics of cellular sensory adaptation in \emph{E. Coli} chemotaxis.

The arguments leading to \eref{eq:refine1} and \eref{eq:refine2} follow from properties of the entropy production rate for Markov processes~\cite{Sartori2014}.
To demonstrate the positivity of \eref{eq:refine1},  we coarse-grain the dynamics by integrating out all the systems that are not neighbors of $X_k$, namely ${\bi X}_{-\Omega_k}$.
For this reduced set of systems, the entropy production rate reduces~\cite{Kawai2007,Parrondo2009,Esposito2012b}, in analogy to \eref{eq:entProd}, to
\begin{equation}
\dot{\tilde S}^k_{\rm i} = \frac{1}{\mu_kT_k}\int \frac{{\tilde J}_k^2(x_k,{\bi x}_{\Omega_k},t)}{p_t(x_k,{\bi x}_{\Omega_k})}\, {\rm d} x_k{\rm d}{\bi x}_{\Omega_k} \ge 0,
\end{equation} 
with the coarse-grained current
\begin{eqnarray}
\eqalign{
{\tilde J}_k&(x_k,{\bi x}_{\Omega_k},t) = \int J_k({\bi x},t)\, {\rm d}{\bi x}_{-\Omega_k} \\
&=\mu_k F_k(x_k,{\bi x}_{\Omega_k},{\bi\lambda}_t)p_t(x_k,{\bi x}_{\Omega_k})-\mu_kT_k\partial_k p_t(x_k,{\bi x}_{\Omega_k}).}
\end{eqnarray}
Now the fact that $F_k$ only depends on the neighbors of $X_k$ becomes crucial: The current  maintains the exact same structure as before in \eref{eq:current}, except now just for $X_k$ and its neighbors ${\bi X}_{\Omega_k}$.
Furthermore, ${\tilde J}_k$ is the only part of the current that enters ${\dot Q}^k$ in \eref{eq:resTot}, that is,
\begin{equation}
{\dot Q}^k =  \int {\tilde J}_k(x_k,{\bi x}_{\Omega_k},t)F_k(x_k,{\bi x}_{\Omega_k},\lambda_t)\, {\rm d}x_k{\rm d}{\bi x}_{\Omega_k}.
\end{equation}
Therefore, we can carry through the exact same steps in deriving \eref{eq:2lawK} to conclude that 
\begin{eqnarray}\label{eq:entNeigh}
\dot{{\tilde S}^k_{\rm i}}=d_tS(X_k)+\frac{{\dot Q}^k}{T_k}-{\dot I}^k(X_k;{\bi X}_{\Omega_k})\ge 0.
\end{eqnarray}
On the other hand, \eref{eq:refine2} follows almost directly from its definition
\begin{eqnarray}
{\dot I}^k(X_k;{\bi X}_{-\Omega_k}|{\bi X}_{\Omega_k})
&= \int J_k({\bi x},t)\partial_k \ln \left[\frac{p_t(x_k,{\bi x}_{-\Omega_k}|{\bi x}_{\Omega_k})}{p_t(x_k|{\bi x}_{\Omega_k})p_t({\bi x}_{-\Omega_k}|{\bi x}_{\Omega_k})}\right]\, {\rm d} {\bi x} \\
& = \int p_t({\bi x}_{-k}){\dot D}^k[p_t(x_k|{\bi x}_{-k})||p_t(x_k|{\bi x}_{\Omega_k})]\,{\rm d} {\bi x}_{-k},
\end{eqnarray}
where we have identified ${\dot D}^k$ as the flow of a relative entropy  [$D(f||g)=\int f(x)\ln [f(x)/g(x)]{\rm d} x$] due to the fluctuations of just $X_k$.
Since the evolution equation at time $t$ with just $X_k$'s dynamics, \emph{i.e.}, $d_tp_t({\bi x})=-\partial_kJ_k({\bi x},t)$, is a Markovian master equation, it will always generate a decrease in a relative entropy, ${\dot D}^k\le 0$~\cite{Sagawa2013}, and \eref{eq:refine2} follows.

\section{Connection to transfer entropy rate}

Having introduced the information flow for multipartite systems, it is important to place this result in the context of the other ways information can be included in thermodynamics.
In particular, Ito and Sagawa have shown that for many interacting systems the transfer entropy rate can also bound the thermodynamics~\cite{Ito2013}.
Namely, the entropy production in the $k$-th system is bounded as
\begin{equation}\label{eq:transIneq}
\sigma^k=d_t{ S}(X_k)+\frac{{\dot Q}^k}{T_k}\ge -{\dot I}(X_k\to{\bi X}_{\Omega_k}),
\end{equation}
where ${\dot I}(X_k\to{\bi X}_{\Omega_k})\ge 0$ is the transfer entropy rate from $X_k$ to its neighbors~\cite{Schreiber2000}.
\Eref{eq:transIneq} codifies how the transfer entropy is a resource like the information flow, allowing for $\sigma^k<0$.
In~\cite{Ito2013}, the transfer entropy rate is explicitly defined for discrete-time dynamics, which is advantageous here because this discrete-time formalism will facilitate our connection with the information flow.
To this end, we consider a trajectory of our systems during a time interval of length $T$.
Time is discretized by breaking it up into $M$ intervals of length $\Delta t = T/M$, and we label each discrete time as $t_\alpha = \alpha \Delta t$, for $\alpha=0,\dots, M$. 
The systems' position at $t_\alpha$ is denoted as ${\bi X}(t_\alpha)=\hat{\bi X}(\alpha)$ and a trajectory up to that time is $\hat{\bi X}_0^\alpha=\{\hat{\bi X}(\beta)\}_{\beta=0}^{\alpha}$.
With this setup the transfer entropy in the $\alpha$ step is given as the mutual information~\cite{Touchette2004,Ito2013}
\begin{equation}\label{eq:trans}
\eqalign{
\fl {\hat I}(X_k\to {\bi X}_{\Omega_k}) &= I[{\hat X}_k(\alpha-1); \hat{\bi X}_{\Omega_k}(\alpha)|(\hat{\bi X}_{\Omega_k})_0^{\alpha-1}]\\
&= \int p[{\hat x}_k(\alpha-1),(\hat{\bi x}_{\Omega_k})_0^\alpha]\ln\frac{p[\hat{\bi x}_{\Omega_k}(\alpha)|\hat{x}_k(\alpha-1),\hat{\bi x}_{\Omega_k}({\alpha-1})]}{p[\hat{\bi x}_{\Omega_k}(\alpha)|(\hat{\bi x}_{\Omega_k})_0^{\alpha-1}]}\, {\rm d}x_k{\rm d}(\hat{\bi x}_{\Omega_k})_0^\alpha.
}
\end{equation}
Then the continuous-time transfer entropy rate is obtained in the limit $\Delta t\to 0$ as ${\dot I}(X_k\to{\bi X}_{\Omega_k})=\lim_{\Delta t\to 0}{\hat I}(X_k\to {\bi X}_{\Omega_k})/\Delta t$.
The transfer entropy is an information-theoretic measure of how much the dynamics of ${\bi X}_{\Omega_k}$ are perturbed by $X_k$.
In this way it offers a measure of the amount of information transferred from $X_k$ to its neighbors.

We will show that the transfer entropy rate gives an upper bound on the information flow as
\begin{equation}\label{eq:transBound}
{\dot I}(X_k\to{\bi X}_{\Omega_k}) \ge \sum_{l\in\Omega_k}{\dot I}^l(X_k;{\bi X}_{\Omega_k}), 
\end{equation}
and thus measures the maximum amount of information that can flow into the neighbors of $X_k$.
The thermodynamic implications of the bound in \eref{eq:transBound} are most strikingly demonstrated in the steady state where all total time derivatives are zero, like $d_tS(X_k)=0$, and are summarized by the series of inequalities
\begin{equation}\label{eq:inequalities}
-\frac{{\dot Q}^k}{T_k} \le -{\dot I}^k(X_k;{\bi X}_{\Omega_k}) =  \sum_{l\in\Omega_k}{\dot I}^l(X_k;{\bi X}_{\Omega_k})\le {\dot I}(X_k\to{\bi X}_{\Omega_k}),
\end{equation}
where the middle equality follows from the fact that the total-time derivative of the mutual information in the steady state is zero: $d_tI = \sum_l{\dot I^l}=0$.
\Eref{eq:inequalities} shows how the amount of heat that can be pulled out of the $k$-reservoir  ${\dot Q}^k$  (as work) is bounded by the information flow due to system $k$, which is ultimately bounded by the transfer entropy rate to $X_k$'s neighbors.
A similar analysis for two interacting systems was presented in~\cite{Horowitz2014b}, and there it was shown that the upper bound on the information flow can be achieved when the auxiliary system implements the Kalman-Bucy filter~\cite{Astrom}.

To demonstrate~\eref{eq:transBound}, we follow the arguments of~\cite{Hartich2014} for two systems.
We start by employing the definition of the conditional probability to rewrite the transfer entropy \eref{eq:trans} as
\begin{equation}
 {\hat I}(X_k\to {\bi X}_{\Omega_k}) = \left\langle \ln\frac{p[\hat{\bi x}_{\Omega_k}(\alpha)|\hat{x}_k(\alpha-1),\hat{\bi x}_{\Omega_k}({\alpha-1})]p[(\hat{\bi x}_{\Omega_k})_0^{\alpha-1}]}{p[(\hat{\bi x}_{\Omega_k})_0^{\alpha}]}\right\rangle,
\end{equation}
where angled brackets denote an average over trajectories.
We then apply the log-sum inequality~\cite{Cover} to $({\bi X}_{\Omega_k})_0^{\alpha-2}$:
\begin{eqnarray}
\fl {\hat I}(X_k\to {\bi X}_{\Omega_k}) &\ge \left\langle\ln\frac{p[\hat{\bi x}_{\Omega_k}(\alpha)|\hat{x}_k(\alpha-1),\hat{\bi x}_{\Omega_k}({\alpha-1})]}{p[{\hat{\bi x}}_{\Omega_k}(\alpha)|\hat{\bi x}_{\Omega_k}(\alpha-1)]}\right\rangle\\
& = \left\langle \ln\frac{p[\hat{\bi x}_{\Omega_k}(\alpha),\hat{\bi x}_{\Omega_k}({\alpha-1})|\hat{x}_k(\alpha-1)]}{p[{\hat{\bi x}}_{\Omega_k}(\alpha),\hat{\bi x}_{\Omega_k}(\alpha-1)]}\right\rangle - \left\langle \ln\frac{p[\hat{\bi x}_{\Omega_k}({\alpha-1})|\hat{x}_k(\alpha-1)]}{p[\hat{\bi x}_{\Omega_k}(\alpha-1)]}\right\rangle.
\end{eqnarray}
One last application of the log-sum inequality on $\hat{\bi X}_{\Omega_k}(\alpha-1)$ in the first term brings us to 
\begin{equation}\label{eq:transInequal2}
\fl {\hat I}(X_k\to {\bi X}_{\Omega_k})  \ge \left\langle \ln\frac{p[\hat{\bi x}_{\Omega_k}(\alpha)|\hat{x}_k(\alpha-1)]}{p[{\hat{\bi x}}_{\Omega_k}(\alpha)]}\right\rangle - \left\langle \ln\frac{p[\hat{\bi x}_{\Omega_k}({\alpha-1})|\hat{x}_k(\alpha-1)]}{p[\hat{\bi x}_{\Omega_k}(\alpha-1)]}\right\rangle.
\end{equation}
The right-hand-side is now the discrete-time definition of the information flow.
We can see this by expanding the $p[\hat{\bi x}_{\Omega_k}(\alpha),\hat{x}_k(\alpha-1)]$ hidden under the brackets in the first term to order $\Delta t$ using the Fokker-Planck equation \eref{eq:fp}.
Noting that it is only the neighbors that progress in time in \eref{eq:transInequal2}, we have
\begin{equation}
\fl p[\hat{\bi x}_{\Omega_k}(\alpha),\hat{x}_k(\alpha-1)]\approx p[\hat{\bi x}_{\Omega_k}(\alpha-1),\hat{x}_k(\alpha-1)]-\Delta t\sum_{l\in \Omega_k} \partial_l J^l[\hat{\bi x}_{\Omega_k}(\alpha-1),\hat{x}_k(\alpha-1)].
\end{equation}
Substituting this into the first term of \eref{eq:transInequal2} gives our desired result
\begin{eqnarray}
\fl {\hat I}(X_k\to {\bi X}_{\Omega_k})  &\gtrsim \Delta t\sum_{l\in\Omega_k}\int J^l[\hat{\bi x}_{\Omega_k}(\alpha-1),\hat{x}_k(\alpha-1)]\partial_l  \ln\frac{p[\hat{\bi x}_{\Omega_k}(\alpha-1)|\hat{x}_k(\alpha-1)]}{p[{\hat{\bi x}}_{\Omega_k}(\alpha-1)]}\\
&\approx \Delta t\sum_{\l \in \Omega_k}{\dot I}^l(X_k;{\bi X}_{\Omega_k}),
\end{eqnarray}
after taking $\Delta t\to 0$.

\section{Illustrative example}

\subsection{Linear systems}\label{sec:linear}
To illustrate the above discussion, we consider a linear system  with forces
\begin{equation}
{\bf F}({\bi X}) = -{\mathbb A}{\bi X}+{\bi f},
\end{equation}
in terms of a time-independent invertible matrix ${\mathbb A}$ with strictly positive eigenvalues, and constant vector ${\bi f}$.
For such constant coupling, the systems will relax to a Gaussian steady state
\begin{equation}
p_{\rm ss}({\bi X}) =\frac{1}{\sqrt{(2\pi)^N|\Sigma|}} \exp\left[-\frac{1}{2}({\bi X}-{\bi m})\Sigma^{-1}({\bi X}-{\bi m})\right],
\end{equation}
characterized by the means ${\bi m} = {\mathbb A}^{-1}{\bi f}$ and the covariance matrix $\Sigma$, given as the solution of $({\bf\mu}{\mathbb A})\Sigma+\Sigma({\bf \mu}{\mathbb A})^T = {\mathbb D}$, with diagonal matrices ${\bf \mu}$ and ${\mathbb D}$ whose elements are ${\bf \mu}_{kl}=\mu_k\delta_{kl}$ and ${\mathbb D}_{kl}=2\mu_kT_k\delta_{kl}$.

It is then a straightforward exercise in integral calculus to determine the steady-state information flow from \eref{eq:infoFlow},
\begin{equation}\label{eq:gaussInfoFlow}
{\dot I}^k(X_k;{\bi X}_{-k}) = \mu_k\left({\mathbb A}_{kk}-T_k\Sigma^{-1}_{kk}\right),
\end{equation}
after some simplifications exploiting the algebraic equation for $\Sigma$.
For the information flow to $X_k$'s neighbors, we need the covariance matrix of just $X_k$ and its neighbors, $\Sigma^{\Omega_k}$, which is obtained by projecting $\Sigma$ onto the subspace of $X_k$ and $X_{\Omega_k}$.
Then,
\begin{equation}\label{eq:gaussInfoFlowNeigh}
{\dot I}^k(X_k;{\bi X}_{\Omega_k}) = \mu_k\left({\mathbb A}_{kk}-T_k(\Sigma^{\Omega_k})^{-1}_{kk}\right).
\end{equation}
As a result, the information flow between $X_k$ and its non-neighbors given its neighbors is
\begin{equation}
{\dot I}^k(X_k;{\bi X}_{-\Omega_k}|{\bi X}_{\Omega_k}) = -\mu_kT_k\left(\Sigma^{-1}_{kk}-(\Sigma^{\Omega_k})^{-1}_{kk}\right)\le 0.
\end{equation}
Finally, from \eref{eq:resTot} the heat flow into the $k$-th reservoir is
\begin{equation}\label{eq:gaussEnt}
{\dot Q}^k= \mu_k\left[({\mathbb A}\Sigma{\mathbb A}^T)_{kk}-T_k{\mathbb A}_{kk}\right].
\end{equation}

\subsection{Two Maxwell demons}
An intriguing application of the above formulas is to consider a system manipulated by two Maxwell demons.
We take as our system of interest an overdamped Brownian particle $X$ trapped in a harmonic well of strength $k$, with mobility $\mu$ in thermal contact with a thermal reservoir at temperature $T=1$, fixing the units of energy.
Its dynamics are given by the Langevin equation
\begin{equation}
{\dot x}_t = -\mu(kx_t - h_t) +\xi_t,
\end{equation}
with zero-mean Gaussian white noise of covariance $\langle\xi_t\xi_s\rangle = 2\mu\delta(t-s)$.
Here, $h_t$ is an external force, which the two Maxwell demons are going to manipulate using feedback in order to extract work.
For concreteness, we identify the internal energy of the particle as solely due to the harmonic trap:  $u= \frac{1}{2}kx_t^2$.
As a result, we have from stochastic energetics~\cite{Sekimoto} that the rate at which work is extracted by $h_t$ is \begin{equation}\label{eq:stochWork}
\dot{w}_{\rm ext} = - h_t\circ{\rm d} x_t,
\end{equation}
where $\circ$ denotes a Stranovich integral.

Feedback is implemented by a pair of identical Maxwell demons $M_1$ and $M_2$, which we model as two Brownian particles with positions $m_{1,t}$ and $m_{2,t}$.
Each demon continuously monitors the position of $X$ and records those measurements in its own position dynamically through the pair of Langevin equations
\begin{equation}
{\dot m}_{i,t} = -\nu k_d (m_{i,t}-x_t)+\eta_{i,t},
\end{equation} 
for $i=1,2$, with $k_d$ their spring constant, $\nu$ the mobility and $\eta_{i,t}$ uncorrelated zero-mean Gaussian white noises of covariance $\langle \eta_{i,t}\eta_{j,s}\rangle = 2\nu\delta_{ij} \delta(t-s)$, uncorrelated with $\xi_t$.
These interactions lead to the neighbor structure illustrated in \fref{fig:neighbors}.
The information acquired by the demons is fed back as a force $h_t = \alpha_1m_{1,t}+\alpha_2m_{2,t}$ designed to extract an average work from $X$'s thermal fluctuations,
\begin{eqnarray}
\eqalign{
 {\dot W}_{\rm ext} = \langle {\dot w}_{\rm ext}\rangle & = -\int J_X(x,m_1,m_2,t)(\alpha_1m_{1}+\alpha_2m_{2})\, {\rm d}x{\rm d}m_1{\rm d}m_2   \\
&\equiv {\dot W}^{M_1}_{\rm ext}+{\dot W}^{M_2}_{\rm ext},
}
\end{eqnarray}
where we have separated out the work extracted by each demon individually.

\begin{figure}
\centering
\includegraphics[scale=0.49]{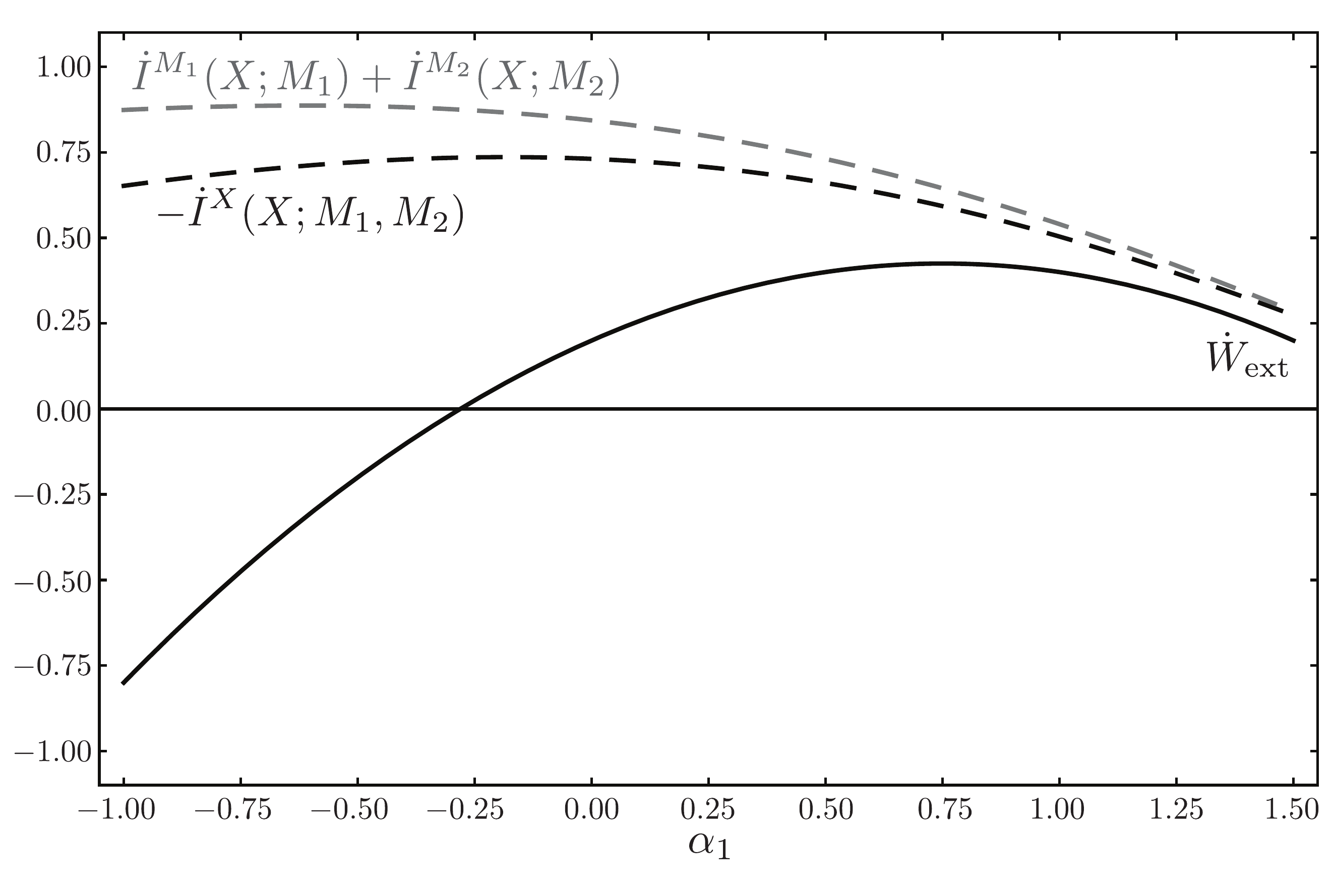}
\caption{Plot of the extracted work ${\dot W}_{\rm ext}$ (solid black) and information flow $-{\dot I}^X(X; M_1,M_2)$ (black dashed) for a Brownian particle controlled by a pair of identical Maxwell demon's as a function of the strength of the feedback force $\alpha_1$ of the first demon. 
For comparison we have included the total amount of information gathered by the pair of demons ${\dot I}^{M_1}(X,M_1)+{\dot I}^{M_2}(X,M_2)$  (gray dashed).  Other parameters are $k=2.5$, $k_d=1.5$, $\mu = 1$, $\nu=2.5$, and $\alpha_2 = 1$.}\label{fig:infoPlot}
\end{figure}

\begin{figure}
\includegraphics[scale=0.335]{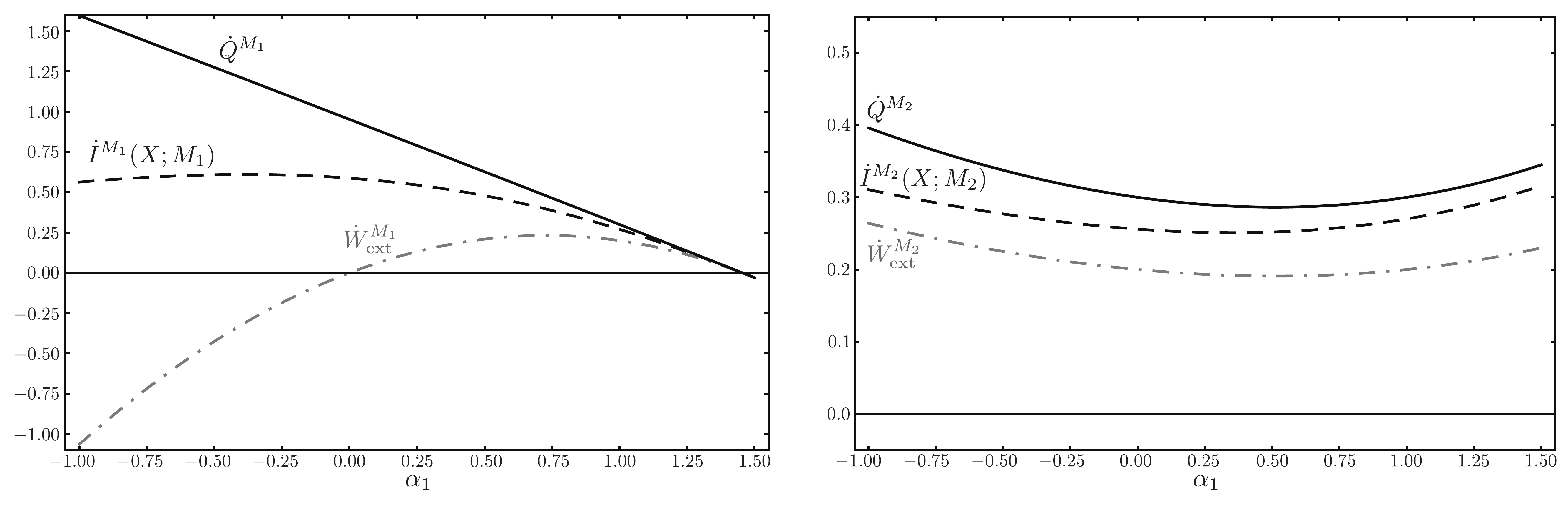}
\caption{(Left) Plot of the heat expended by the first demon ${\dot Q}^{M_1}$ (black) to gather information ${\dot I}^{M_1}(X;M_1)$ (black dashed), which exceeds the work it extracts ${\dot W}_{\rm ext}^{M_1}$ (gray dot-dashed) as a function of the strength of the feedback force $\alpha_1$.  (Right) Same for second demon, $M_2$.
Parameters are $k=2.5$, $k_d=1.5$, $\mu = 1$, $\nu=2.5$, and $\alpha_2 = 1$.}\label{fig:infoPlot2}
\end{figure}

Now, in the steady state the first law of thermodynamics demands that ${\dot W}_{\rm ext}=-{\dot Q}^X$, and the information-flow inequality in \eref{eq:entNeigh} becomes 
\begin{equation}\label{eq:workInfoBound}
{\dot W}_{\rm ext} \le -{\dot I}^X(X;M_1,M_1).
\end{equation}
Thus, the amount of work extracted by the demons is bounded by the amount of information consumed by $X$.
We demonstrate this in \fref{fig:infoPlot} using the formulas of \sref{sec:linear}.
Of course, the ultimate source of this information are the two demons, which must supply heat in accordance with the refined information-flow inequality in \eref{eq:refine1},
\begin{equation}
{\dot Q}^{M_1} \ge  {\dot I}^{M_1}(X;M_1), \qquad {\dot Q}^{M_2} \ge  {\dot I}^{M_2}(X;M_2),
\end{equation}
 as verified in \fref{fig:infoPlot2}. 
 Remarkably, the work extracted by each demon individually is bounded by the information it has: $ {\dot W}_{\rm ext}^{M_i} \le I^{M_{i}}(X;M_{i})$, $i=1,2$. 
 How general this observation is remains to be determined.
 Unfortunately, the demons supply more information than is necessary for work extraction because they operate independently, which we see graphically in \fref{fig:infoPlot} through the inequality ${\dot I}^{M_1}(X,M_1)+{\dot I}^{M_2}(X,M_2)>-{\dot I}^X(X;M_1,M_2)$.
 
\section{Conclusion}

We have introduced a refinement of the second law for many interacting, multipartite systems that each experience independent, uncorrelated noise.
For such a collection of systems, the second law splits into a separate, positive contribution for each system.
The interactions among the systems then manifests itself in this entropy balance by the appearance of an information flow that accounts for how each system learns about the others and utilizes that information.
We then further refined this statement by recognizing that each system only directly interacts with a subset of all the other systems, a group we call its neighbors.
As a result, the information flow can be restricted to a system and just its neighbors, which implies, among other things, that only information shared between a system and its neighbors can be used to perform a useful task.

A careful analysis our results also reveals that our key requirement is that each system experiences independent noise.
We do not need to require that these noises be generated by distinct thermodynamic reservoirs, though we have made this assumption for clarity of presentation.
All we really need is to be able to track the thermodynamic flows from each system into the surroundings.
Thus, for example, our results apply to a collection of small systems immersed in a common thermal reservoir, as long as the perturbations affecting each system from the reservoir are independent.
Still, not all collections of systems where information dynamics are interesting will satisfy our multipartite assumption.
In this direction, a growing amount of research has been focused on information reservoirs~\cite{Mandal2012,Deffner2013,Barato2014c,Barato2013c,Hoppenau2014}, which do not have a multipartite structure.

We further imagine that the analysis presented here will prove helpful in re-investigating the energetics of communication networks as outlined by Landauer~\cite{Landauer1996}, since such networks of communication channels are physically implemented by many interacting, small thermodynamic systems.
Particularly intriguing examples are biochemical sensory networks responsible for transducing signals into and through cells, which can have a multipartite structure~\cite{Barato2014b,Sartori2014,Mehta2012,Lan2012,Ito2014,Smith}.

\ack
This work was supported by the ARO MURI grant W911NF-11-1-0268.


\section*{References}

\bibliography{Feedback.bib,PhysicsTexts.bib}
\bibliographystyle{iopart-num}

\end{document}